\documentclass[journal]{IEEEtran}

\usepackage{setspace}
\usepackage{cite,graphicx,amsmath,amssymb}
\usepackage{subfigure}
\usepackage{fancyhdr}
\usepackage{mdwmath}
\usepackage{mdwtab}
\usepackage{balance}
\usepackage{xcolor}
\usepackage{bm}
\usepackage{amsthm}
\usepackage{threeparttable}
\usepackage{algorithm}
\usepackage{algorithmic}
\usepackage{multirow}
\usepackage{flafter}
\usepackage{makecell}
\usepackage{diagbox}
\usepackage{graphicx,epstopdf}
\usepackage{epsfig}

\newcommand{\tabincell}[2]{\begin{tabular}{@{}#1@{}}#2\end{tabular}}
\providecommand{\url}[1]{#1}
\allowdisplaybreaks
\begin{document}
\title{NOMA for Integrating Sensing and Communications towards 6G: A Multiple Access Perspective}
\author{
Xidong~Mu, Zhaolin~Wang, and Yuanwei~Liu

\thanks{X. Mu, Z. Wang, and Y. Liu are with the School of Electronic Engineering and Computer Science, Queen Mary University of London, London E1 4NS, UK, (email: \{x.mu, zhaolin.wang, yuanwei.liu\}@qmul.ac.uk).}
}
\maketitle
\begin{abstract}
This article focuses on the development of integrated sensing and communications (ISAC) from a multiple access (MA) perspective, where the idea of non-orthogonal multiple access (NOMA) is exploited for harmoniously accommodating the sensing and communication functionalities. We first reveal that the developing trend of ISAC is from \emph{orthogonality} to \emph{non-orthogonality}, and introduce the fundamental models of the downlink and uplink ISAC while identifying the design challenges from the MA perspective. (1) For the downlink ISAC, we propose two novel designs, namely \emph{NOMA-empowered} downlink ISAC and \emph{NOMA-inspired} downlink ISAC to effectively coordinate the inter-user interference and the sensing-to-communication interference, respectively. (2) For the uplink ISAC, we first propose a \emph{pure-NOMA-based} uplink ISAC design, where a fixed communication-to-sensing successive interference cancellation order is employed for distinguishing the mixed sensing-communication signal received over the fully shared radio resources. Then, we propose a general \emph{semi-NOMA-based} uplink ISAC design, which includes the conventional orthogonal multiple access-based and pure-NOMA-based uplink ISAC as special cases, thus being capable of providing flexible resource allocation strategies between sensing and communication. Along each proposed NOMA-ISAC design, numerical results are provided for showing the superiority over conventional ISAC designs.
\end{abstract}

\section{Introduction}

\begin{figure*}[!t]
  \centering
  \includegraphics[width=0.7\textwidth]{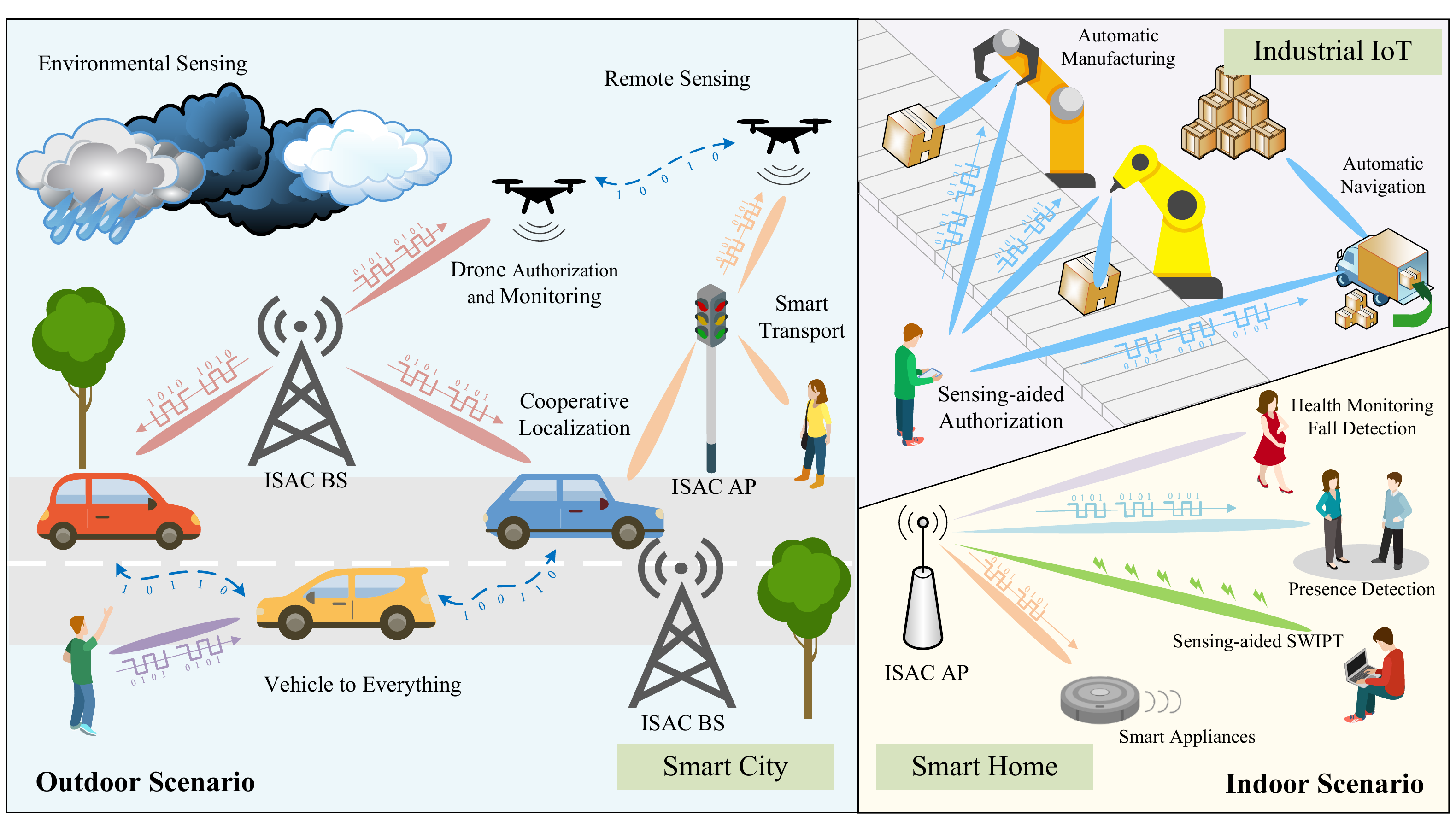}\\
  \caption{Applications of ISAC in future wireless networks.}\label{fig:application}
\end{figure*}

Recently, the concept of integrated sensing and communications (ISAC) has attracted significant attention from both industry and academia \cite{liu2021integrated,zhang2021enabling,tan2021integrated}. This is mainly driven by the unprecedented demands on sensing in next generation wireless networks for supporting various ``intelligence'' enabled applications \cite{latva2020key}. Among the family of sensing techniques, wireless radio sensing has been recognized as a promising sensing approach in 6G. For achieving high-quality and ubiquitous sensing, wireless radio sensing has evolved in some similar directions to wireless communications, including ultra-wideband, high radio frequencies, massive multiple-input multiple-output (MIMO), and ultra-dense networks \cite{liu2021integrated,latva2020key}. This thus opens up new exciting possibilities to integrate sensing and communication functionalities by sharing the same hardware platforms, radio resources, and signal processing pipelines, which motivates the research theme of ISAC. On the one hand, ISAC can enhance the resource efficiency of energy, spectrum, and hardware. For example, ISAC can equip the wireless communication networks with the sensing functionality without the need for significant changes to existing wireless infrastructures, i.e., leading to a relatively low cost. On the other hand, through the deep integration and novel protocol designs, ISAC can further achieve mutualism benefits compared to the single-functional network \cite{liu2021integrated}. Given the aforementioned potential advantages, as shown in Fig. \ref{fig:application}, ISAC can be extensively employed in the future wireless networks, including but not limited to smart cities, industrial Internet-of-things (IoT), and smart homes.

Despite being promising for 6G, the strike of a good performance tradeoff between the two functionalities is a challenging task when designing ISAC \cite{liu2021integrated}. The intrinsic reason is that ISAC may suffer from severe inter-functionality interference due to the hardware platform and radio resource sharing. This calls for the development of efficient interference mitigation and resource management approaches. Reviewing the development road of wireless communications from 1G to 5G, multiple access (MA) techniques have long been one of the most fundamental enablers for efficiently accommodating multiple communication users via appropriate interference control and resource allocation \cite{liu2022evolution}. Following the same path, this article aims to investigate the development of ISAC from the MA perspective with a particular focus on non-orthogonal multiple access (NOMA). By employing the superposition coding (SC) at the transmitter and the successive interference cancellation (SIC) at the receiver, NOMA allows multiple communication users to be well served over the same radio resources, thus significantly enhancing the connectivity
and improving the resource efficiency \cite{liu2022evolution}. The prominent features of NOMA in efficient interference management and flexible resource allocation match well with the requirements encountered in ISAC. Nevertheless, utilizing NOMA in ISAC is not straightforward and requires re-designs. This is because NOMA is currently employed to serve users only for the communication functionality, and there is no answer on how to employ NOMA to support the two functionalities in ISAC. This motivates us to contribute this article.

The organization of this article is as follows. The developing trend of ISAC is discussed with the idea from orthogonality to non-orthogonality, followed by introducing the fundamental models of downlink and uplink ISAC as well as the corresponding design challenges from the MA perspective. Against these challenges, several novel NOMA-ISAC designs are proposed for both the downlink and uplink ISAC, where numerical results are provided for showing the effectiveness. Finally, conclusions are drawn.

\section{ISAC: Trends, Models, and Challenges}

\subsection{From Orthogonal-ISAC to Non-Orthogonal-ISAC}
The main idea of ISAC is to facilitate both the sensing and communication functionalities in harmony via the same hardware platform. This is motivated by the fact that sensing and communication share some philosophy in common, i.e., relying on the exploitation of radio waves. Communication exploits the radio wave to convey information bits from the transmitter to the receiver, while sensing exploits the radio wave echo reflected by the target to analyze and obtain the desired sensing-related parameters (e.g., the location, speed, and shape of the target) \cite{zhang2021overview}. The ultimate goal of ISAC is to go beyond the two separate functionalities and facilitate promising interplay between them for achieving mutualism benefits, namely communication-assisted sensing and sensing-assisted communication~\cite{liu2021integrated}. Nevertheless, on the road to ISAC, one of the most fundamental issues is how to effectively coordinate the mutual interference between sensing and communication. One straight solution is to accommodate the sensing and communication functionalities in the allotted \emph{orthogonal} radio resources (e.g., in time-/frequency-/space-/code domains) \cite{zhang2022time,shi2017power,chen2021code}, which can be generally termed as orthogonal-ISAC. Despite having a low implementation complexity and an interference-free sensing/communication process, the resultant spectrum- and energy-efficiency of orthogonal-ISAC would be low since the radio resources have to be strictly orthogonally used. To improve resource efficiency and promote the integration level, it is necessary to develop ISAC by sharing the same radio resources. In this case, since the sensing and communication functionalities have to be facilitated in a \emph{non-orthogonal} manner with properly designed inter-functionality interference mitigation approaches, we term this as non-orthogonal-ISAC. Besides the improved resource efficiency, non-orthogonal-ISAC exhibits potential advantages, such as high compatibility and flexibility. For instance, non-orthogonal-ISAC enables communication (sensing) to be seamlessly integrated with sensing (communication) in the spectrum resource which has been already occupied by sensing (communication) in history~\cite{liu2021integrated}. However, in order to fully reap the benefits provided by non-orthogonal-ISAC, efficient methods have to be developed for mitigating the inter-functionality interference. In the following, we introduce the fundamental models of downlink and uplink ISAC and identify the corresponding design challenge in each model if employing non-orthogonal-ISAC. Unless stated otherwise, we use ``ISAC'' to refer to ``non-orthogonal-ISAC'' in the remaining context.

\subsection{Downlink ISAC}
\begin{figure*}[!t]
\centering
\subfigure[]{\label{DL}
\includegraphics[width= 2.5in]{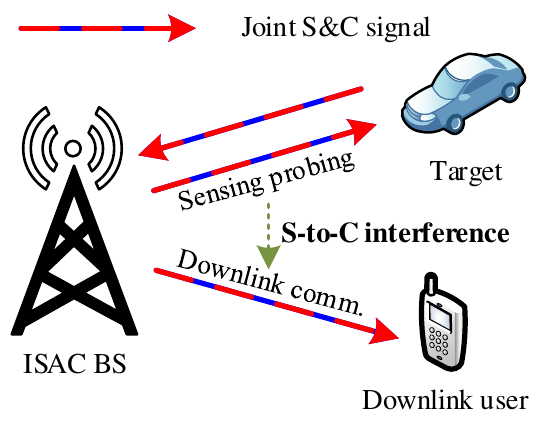}}
\subfigure[]{\label{UL}
\includegraphics[width= 2.5in]{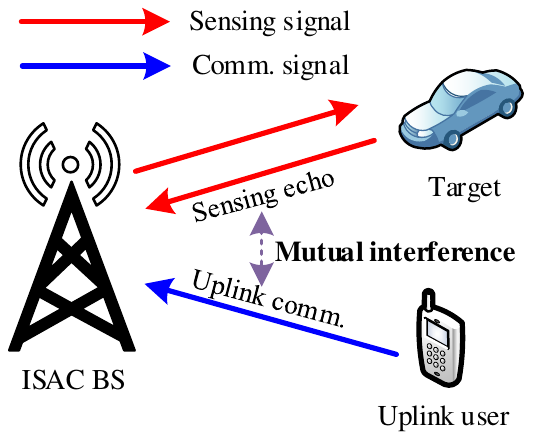}}
\caption{Illustration of (a) a downlink ISAC model and (b) an uplink ISAC model. In each model, there is one ISAC BS, one sensing target, and one communication user.}
\end{figure*}

In order to better introduce the concept of ISAC, we consider a simple model, which includes one ISAC base station (BS), one sensing target, and one communication user. A typical downlink ISAC is illustrated in Fig. \ref{DL}. On the one hand, the ISAC BS sends the sensing probing signal, which will be reflected by the target and returned back as the sensing echo. Thus, ISAC BS is able to estimate the relevant sensing parameters by analyzing the received sensing echo. On the other hand, the ISAC BS simultaneously sends the communication signal to the communication user. From the MA perspective, it can be observed that both the sensing and communication signals are transmitted by the ISAC BS, and the sensing parameter estimation and the information bit decoding are carried out at two different destinations (i.e., the ISAC BS for sensing and the downlink user for communication). Therefore, we classify this type of model as the downlink ISAC.

Recall the fact that sensing can be only carried out at the ISAC BS until receiving the echo of the transmit sensing waveform reflected by the target. In other words, there is no information contained in the transmit sensing waveform at first as that in the transmit communication waveform. Inspired by this observation, the key task in the downlink ISAC is the design of the joint S\&C waveforms, which are capable of achieving both sensing and communication \cite{hua2021optimal,liu2020joint,liu2018toward}. Since the ISAC BS prior knows the communication information and the analyze of the sensing echo also does not focus on the information bits modulated on the joint S\&C waveforms, there is in general no communication-to-sensing interference in the downlink ISAC. However, the reverse is not true. This is because for well achieving sensing, the number of required joint S\&C waveforms may be larger than the number of downlink communication streams~\cite{liu2020joint,hua2021optimal}. As a result, some of joint waveforms (termed as additional sensing waveforms) will lead to \emph{sensing-to-communication interference} as shown in Fig. \ref{DL}. How to mitigate the sensing-to-communication interference is the major design challenge in the downlink ISAC, which will be discussed with NOMA in Section III.

\subsection{Uplink ISAC}

We continue to introduce the fundamental model of the uplink ISAC, as shown in Fig. \ref{UL}. In this case, the entire procedure of sensing remains unchanged, while the communication signal is uploaded by the uplink user to the ISAC BS. From the MA perspective, the sensing and communication signals are respectively transmitted by the ISAC BS and the uplink user, while the sensing parameter estimation and the information bit decoding are carried out at the same destination (ISAC BS). We classify this type of model as the uplink ISAC. It can be observed that since the sensing and communication signals are separately transmitted, there is no need to design the joint S\&C waveforms in the uplink ISAC.

In the uplink ISAC, the ISAC BS has to analyze the sensing echo for obtaining the sensing results and decoding the communication signal for recovering the information message. Both of them are not prior known by the ISAC BS. When the ISAC BS processes the received mixed sensing and communication signals, the major design challenge in the uplink ISAC is how to mitigate the \emph{mutual interference} between sensing and communication, as shown in Fig. \ref{UL}. The possible solutions with NOMA will be discussed in Section IV.

\section{NOMA-enhanced Downlink ISAC}
\begin{figure*}[!t]
\centering
\subfigure[]{\label{fig:model_noma_empowered}
\includegraphics[width= 2.1in, height= 1.6in]{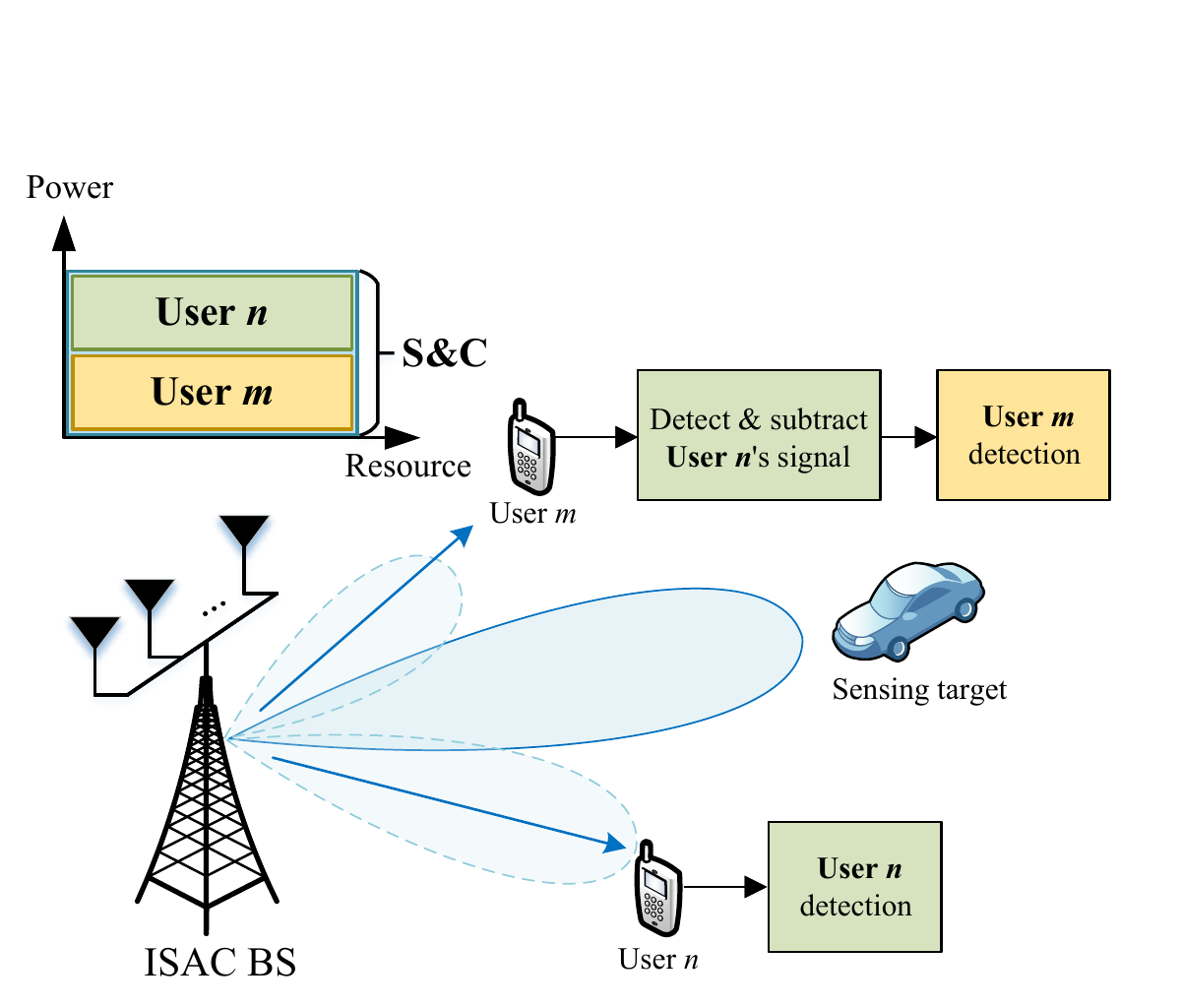}}
\subfigure[]{\label{fig:tradeoff_noma_empowered}
\includegraphics[width= 2in, height= 1.6in]{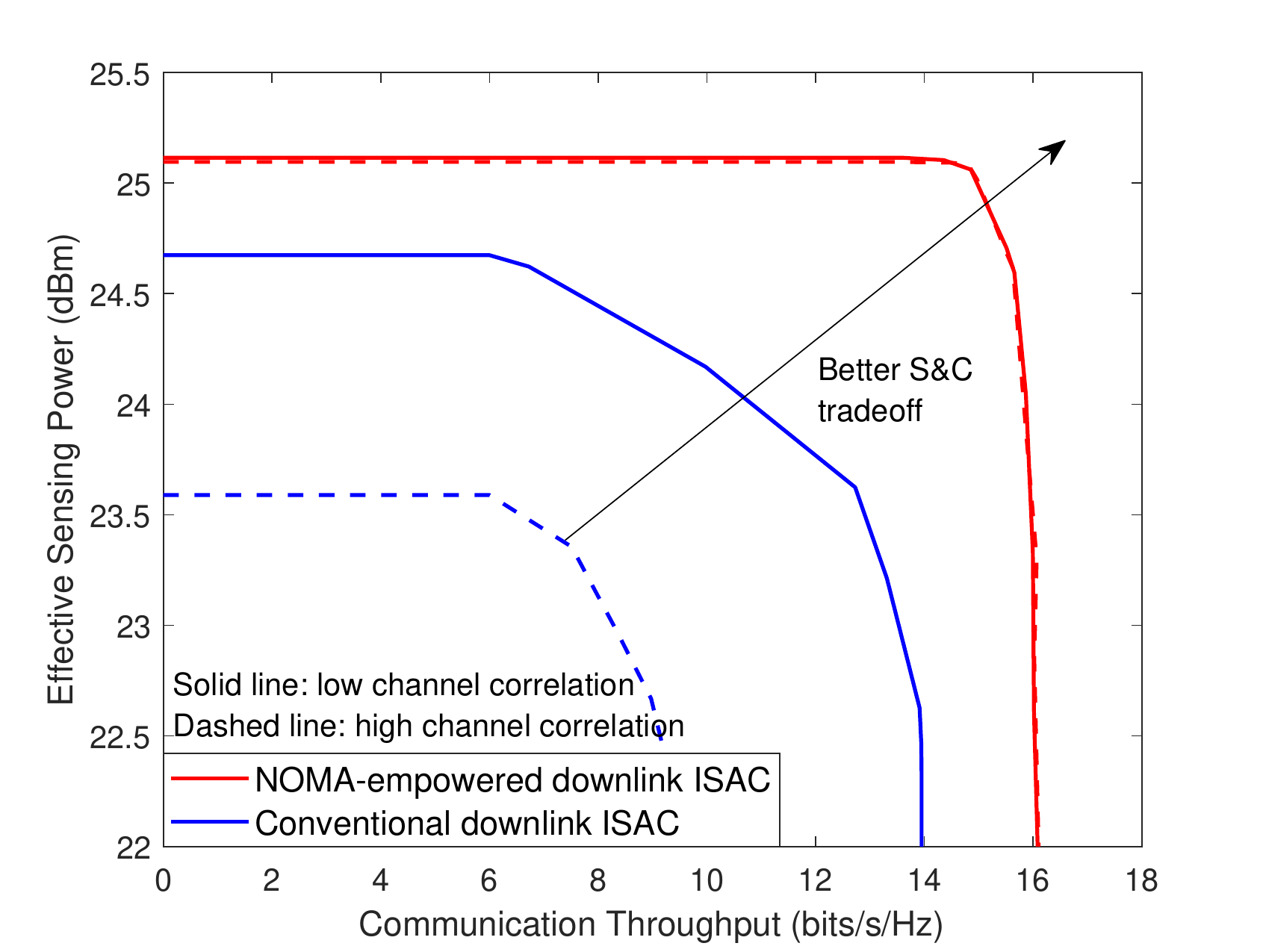}}
\subfigure[]{\label{fig:tradeoff_noma_empowered_2}
\includegraphics[width= 2in, height= 1.6in]{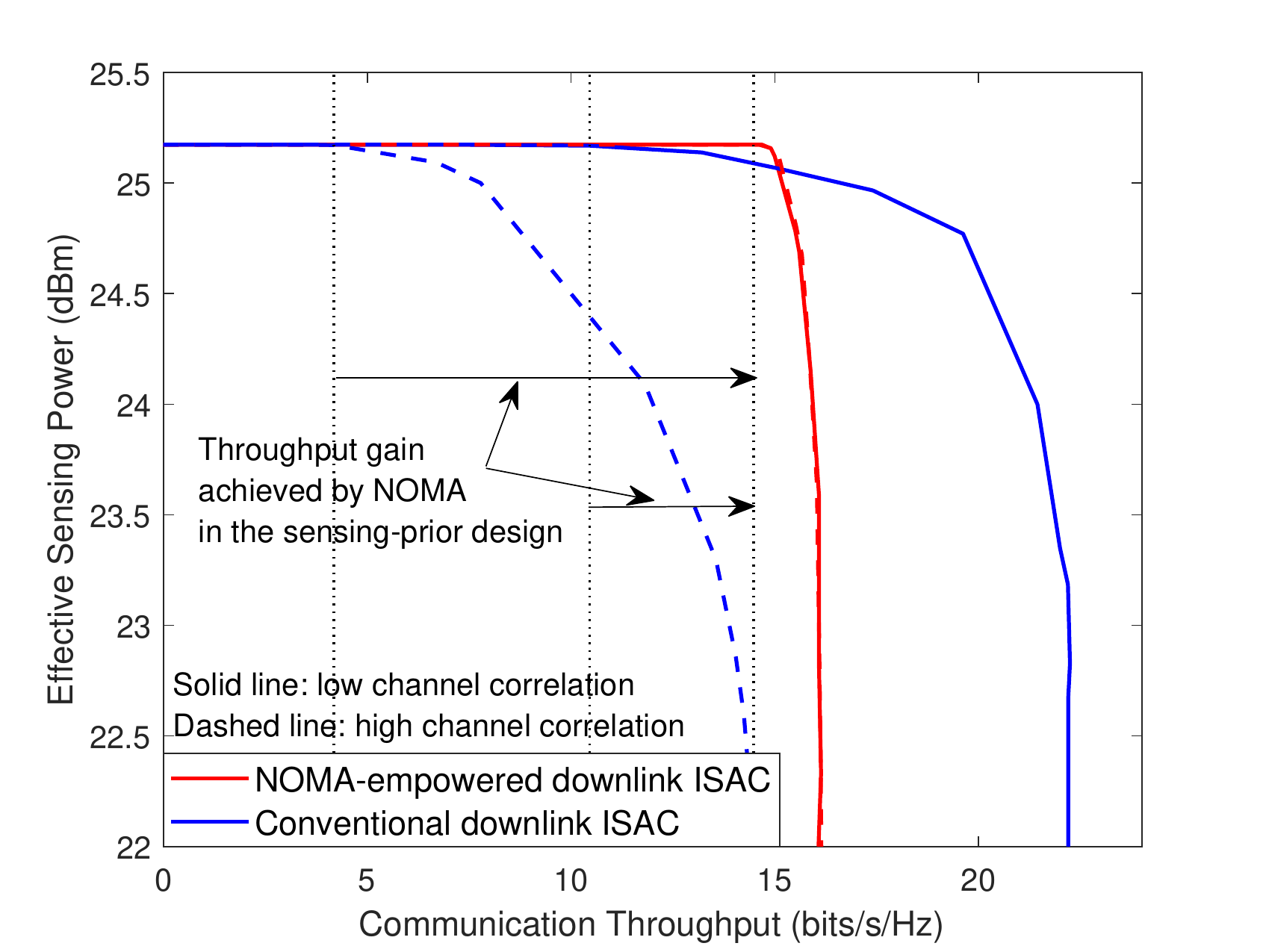}}
\caption{Illustration of (a) the NOMA-empowered downlink ISAC design and the achieved sensing-versus-communication tradeoff in (b) the overloaded regime and (c) the underloaded regime.
  The system parameter settings can be found in \cite[Section IV]{wang2022noma}}
\end{figure*}
On the one hand, as discussed in Section II.B, one main task in the downlink ISAC is to design the joint S\&C waveforms for combating the potential sensing-to-communication interference. On the other hand, the typical inter-user communication interference management will become more challenging in the downlink ISAC since the support of sensing may introduce additional requirements on the joint S\&C waveform design. To this end, we explore the possible applications of NOMA to enhance the performance of downlink ISAC. In particular, we propose two novel designs, namely 1) NOMA-empowered downlink ISAC and 2) NOMA-inspired downlink ISAC, for efficiently mitigate the typical inter-user interference and the new sensing-to-communication interference, thus achieving a good sensing-versus-communication tradeoff. The key difference between the two designs is whether NOMA is employed within the communication functionality or between the two functionalities in ISAC.

\subsection{NOMA-empowered Downlink ISAC Design}

In the NOMA-empowered downlink ISAC design, as shown in Fig. \ref{fig:model_noma_empowered}, the SC and SIC techniques are invoked for transmitting and detecting the communication signal for each user. Moreover, the superimposed communication signals are also exploited for target sensing. Therefore, the number of joint S\&C waveforms is equal to the number of users, thus leading to no additional sensing-to-communication interference. This can be regarded as a straightforward extension of employing NOMA from the conventional communication system to the ISAC system, i.e., the communication functionality of ISAC is \emph{empowered} by NOMA. Consequently, the NOMA-empowered downlink ISAC design inherits the key feature of conventional NOMA, namely the additional DoFs provided by SIC for inter-user interference cancellation. In conventional ISAC~\cite{liu2018toward}, the joint S\&C waveform designs for mitigating the inter-user interference merely rely on the spatial DoFs. As a result, it may suffer from severe performance degradation when the spatial DoFs are deficient. Fortunately, this problem can be addressed by the proposed NOMA-empowered downlink ISAC with the aid of SIC. In the following, to fully unveil the advantages of the proposed NOMA-empowered downlink ISAC design, we discuss it in both the overloaded and underloaded regimes.

In the overloaded regime, it is impossible to allocate at least one spatial DoF per communication user even without sensing. Therefore, the corresponding inter-user interference cannot be effectively suppressed in the spatial domain. In this case, the lack of efficient inter-user interference mitigation not only causes degraded communication performance but also further damages sensing performance given the limited DoFs in joint S\&C waveform designs. As a remedy, NOMA-empowered downlink ISAC provides extra DoFs for inter-user interference mitigation via SIC and thus achieves enhanced sensing and communication performance than the conventional downlink ISAC designs. To show this benefit, Fig. \ref{fig:tradeoff_noma_empowered} compares the sensing-versus-communication tradeoff region achieved by the proposed NOMA-empowered downlink ISAC and the conventional downlink ISAC in the overloaded regime \cite{wang2022noma}. As can be observed, NOMA-empowered downlink ISAC is capable of significantly enlarging the sensing-versus-communication tradeoff region compared with the conventional downlink ISAC~\cite{liu2018toward}. Moreover, it also shows that the users' channel correlation almost has no effect on the performance achieved by NOMA-empowered downlink ISAC, i.e., yielding superior robustness.

In contrast to the overloaded regime where the spatial DoFs are always insufficient, the deficiency of spatial DoFs in the underloaded regime may also occur in following two cases. On the one hand, the correlated users' channels will directly cause the deficiency of spatial DoFs in the underloaded regime. On the other hand, in the sensing-prior design of ISAC, the joint S\&C waveforms transmitted by the ISAC BS need to be constructed in some specific beam patterns for ensuring high-quality sensing, which might limit the spatial DoFs available for inter-user interference mitigation. To better illustrate this, in Fig. \ref{fig:tradeoff_noma_empowered_2}, we compare the performance achieved by the proposed NOMA-empowered downlink ISAC and the conventional downlink ISAC in the aforementioned two cases \cite{wang2022noma}. As expected, the proposed NOMA-empowered downlink ISAC is superior when the users' channel correlation is high. Moreover, in the sensing-prior design where the maximum sensing performance has to be achieved, NOMA-empowered downlink ISAC can achieve a higher communication throughput than the conventional downlink ISAC regardless of the users' channel correlation. Nevertheless, it can be also observed that NOMA-empowered downlink ISAC is not effective in the underloaded case with low users' channel correlations and relaxed sensing requirements. This is also expected since the available spatial DoFs in these cases are sufficient and the SIC operation is redundant~\cite{liu2022evolution}. However, given the rapidly increasing number of connected devices, future wireless networks will be more overloaded than previous and current wireless networks. It indicates that the proposed NOMA-empowered downlink ISAC design will play an important role.

\subsection{NOMA-inspired Downlink ISAC Design}

\begin{figure*}[!t]
\centering
\subfigure[]{\label{fig:model_noma_inspired}
\includegraphics[width= 3in]{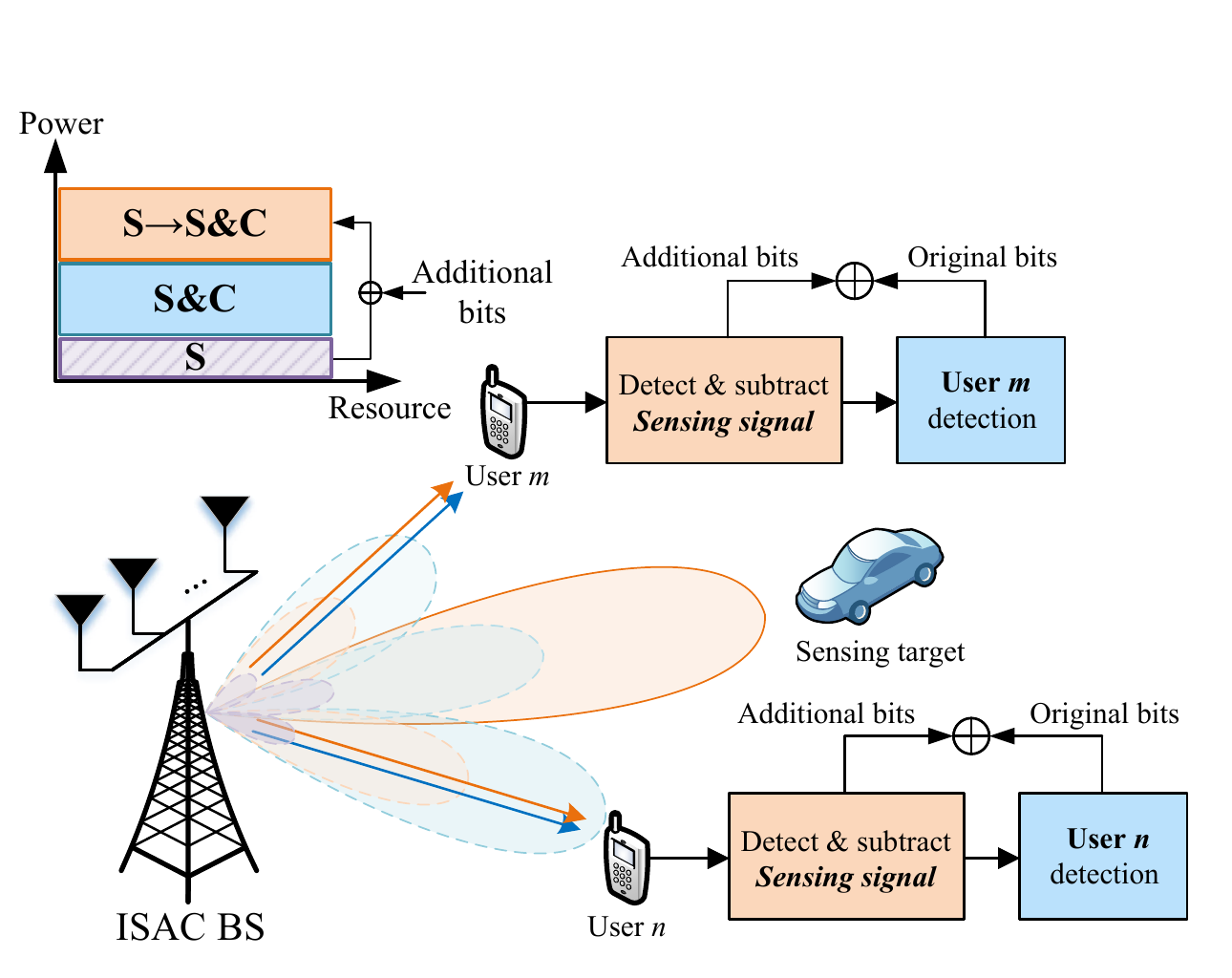}}
\subfigure[]{\label{fig:tradeoff_noma_inspired}
\includegraphics[width= 3in]{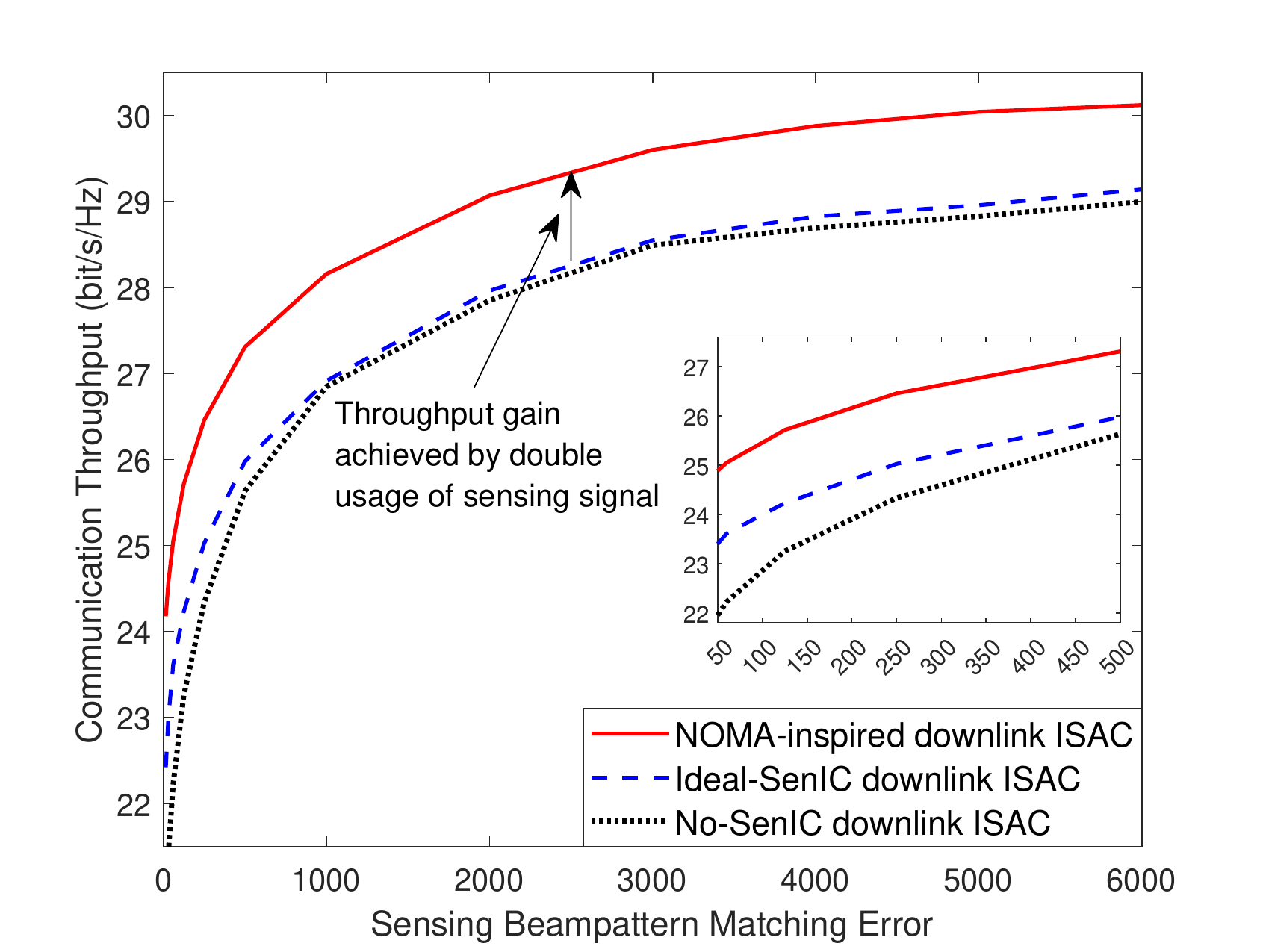}}
\caption{Illustration of (a) the NOMA-inspired downlink ISAC design and (b) the sensing-versus-communication tradeoff region achieved by different downlink ISAC designs.
  Here, ``Ideal SenIC'' refers to ideally removing sensing interference at the user \cite{hua2021optimal} and ``No SenIC'' refers to not removing the sensing interference \cite{liu2020joint}. The system parameter settings can be found in \cite[Section IV]{wang2022inspired}.}
\end{figure*}
\begin{table*}[t!]
  \caption{Summary of the Proposed NOMA enhanced Downlink ISAC Designs}
  \scriptsize
  \centering
  \begin{tabular}{|l|l|l|c|}
      \hline
                                           & \textbf{Characteristic}                             & \textbf{Advantages}       & \textbf{Ref.}                                   \\ \hline
      NOMA-empowered downlink ISAC                  & NOMA within C.                  & \tabincell{l}{1. Mitigating \emph{inter-user interference}. \\ 2. Accommodating sensing-prior design.}       & \cite{wang2022noma}                                  \\ \hline
      NOMA-inspired downlink ISAC                   & NOMA between S\&C.     & \tabincell{l}{1. Mitigating \emph{sensing-to-communication interference}. \\ 2. Fully exploiting the additional sensing signal for communication.}    & \cite{wang2022inspired}                                    \\ \hline
  \end{tabular}
  \label{table:comparison}
\end{table*}

In the above proposed NOMA-empowered downlink ISAC design, we directly apply NOMA for addressing the potential stringent inter-user interference. In this subsection, we handle the emerging sensing-to-interference issue in the downlink ISAC. As pointed out by \cite{liu2020joint,hua2021optimal}, additional sensing waveforms are usually essential in downlink ISAC for achieving high-quality sensing, but cause harmful interference to communication. To overcome this obstacle in the downlink ISAC, we propose a NOMA-inspired downlink ISAC design for providing a practical sensing interference mitigation approach, which can be further modified to also benefit communication. As shown in Fig. \ref{fig:model_noma_inspired}, the main idea of NOMA-inspired downlink ISAC is to allow part or all of the additional sensing waveforms to convey information bits, thus becoming additional joint S\&C waveforms. Consequently, these additional joint S\&C waveforms can be regarded to be intended for \emph{virtual} communication users, who are paired with \emph{real} communication users via NOMA. At each real communication user, the signal of the virtual communication users (i.e., the additional joint S\&C waveforms) will be first detected and removed via SIC, thus mitigating the sensing-to-communication interference in the downlink ISAC. Generally speaking, NOMA-inspired downlink ISAC can be viewed as employing NOMA among the sensing and communication functionalities for interference mitigation. As a further advance, the information bits conveyed in the additional joint S\&C waveforms can be also made useful to communication users. For example, the additional joint S\&C waveforms in \cite{wang2022inspired} are employed to convey multicast messages intended for all communication users, thus benefiting both sensing and communication, namely double benefits.

Fig. \ref{fig:tradeoff_noma_inspired} compares the achieved sensing-versus-communication performance of the proposed NOMA-inspired downlink ISAC design, where the additional sensing signal is exploited for additional multicast transmission. For comparison, two baseline downlink ISAC designs are considered, where the additional sensing signals are assumed to be ideally removed~\cite{hua2021optimal} and not removed~\cite{liu2020joint}. It can be observed that the proposed NOMA-inspired downlink ISAC design achieves a significant performance gain due to the further exploitation of additional sensing signals. This result confirms the effectiveness of the proposed NOMA-inspired downlink ISAC design.

In Table \ref{table:comparison}, we summarize the respective characteristics and advantages of the above two NOMA-ISAC designs in downlink. It can be observed that the two designs individually focus on the mitigation of inter-user interference and sensing-to-communication interference. However, in practice, the two types of interference will simultaneously exist in the downlink ISAC. In this context, a unified NOMA-enhanced downlink ISAC design needs to be developed, which is capable of adaptively carrying out different types of interference mitigation. This is an interesting future research topic.

\section{NOMA-based Uplink ISAC}

\begin{figure*}[!t]
\centering
\subfigure[]{\label{fig:UL model}
\includegraphics[width= 3.3in]{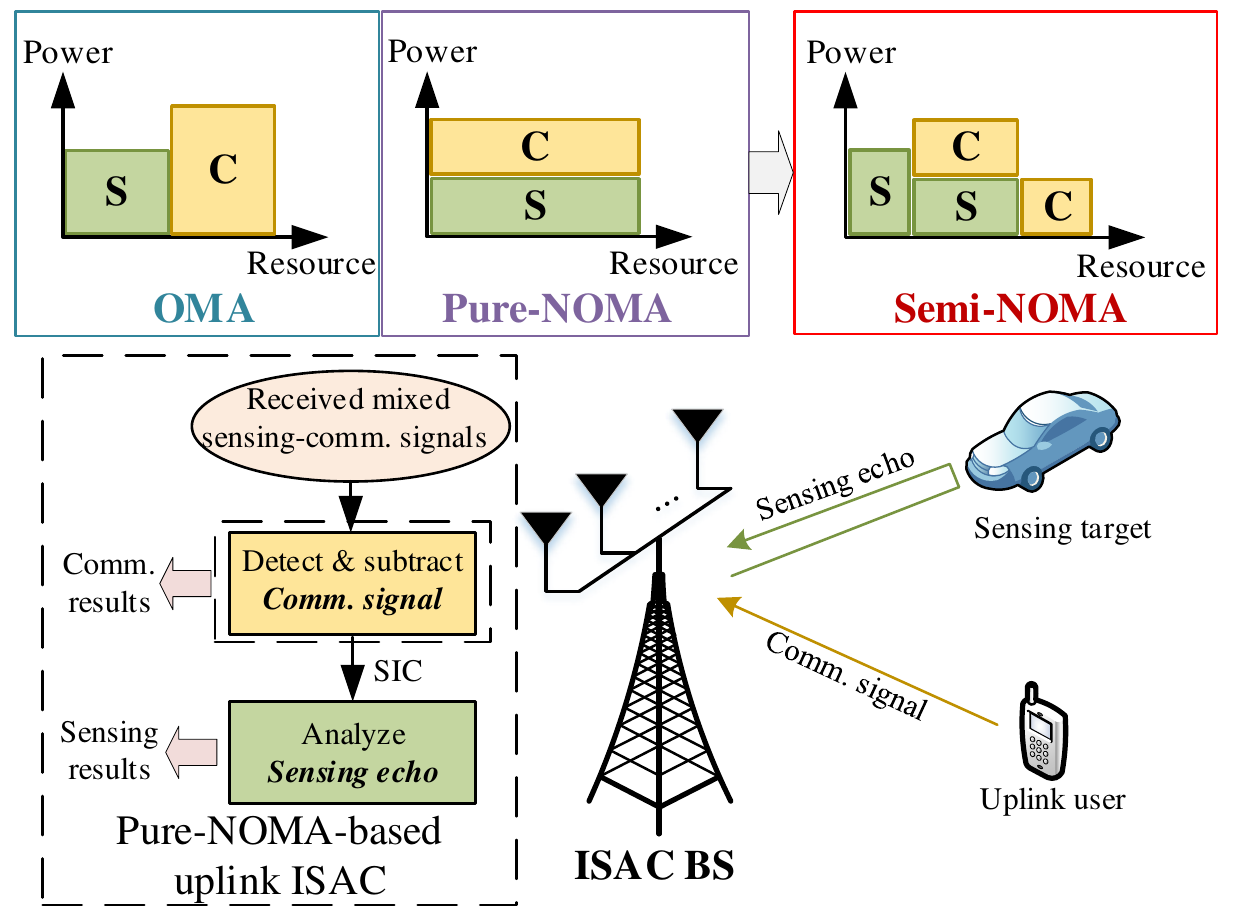}}
\subfigure[]{\label{semi_result}
\includegraphics[width= 2.8in]{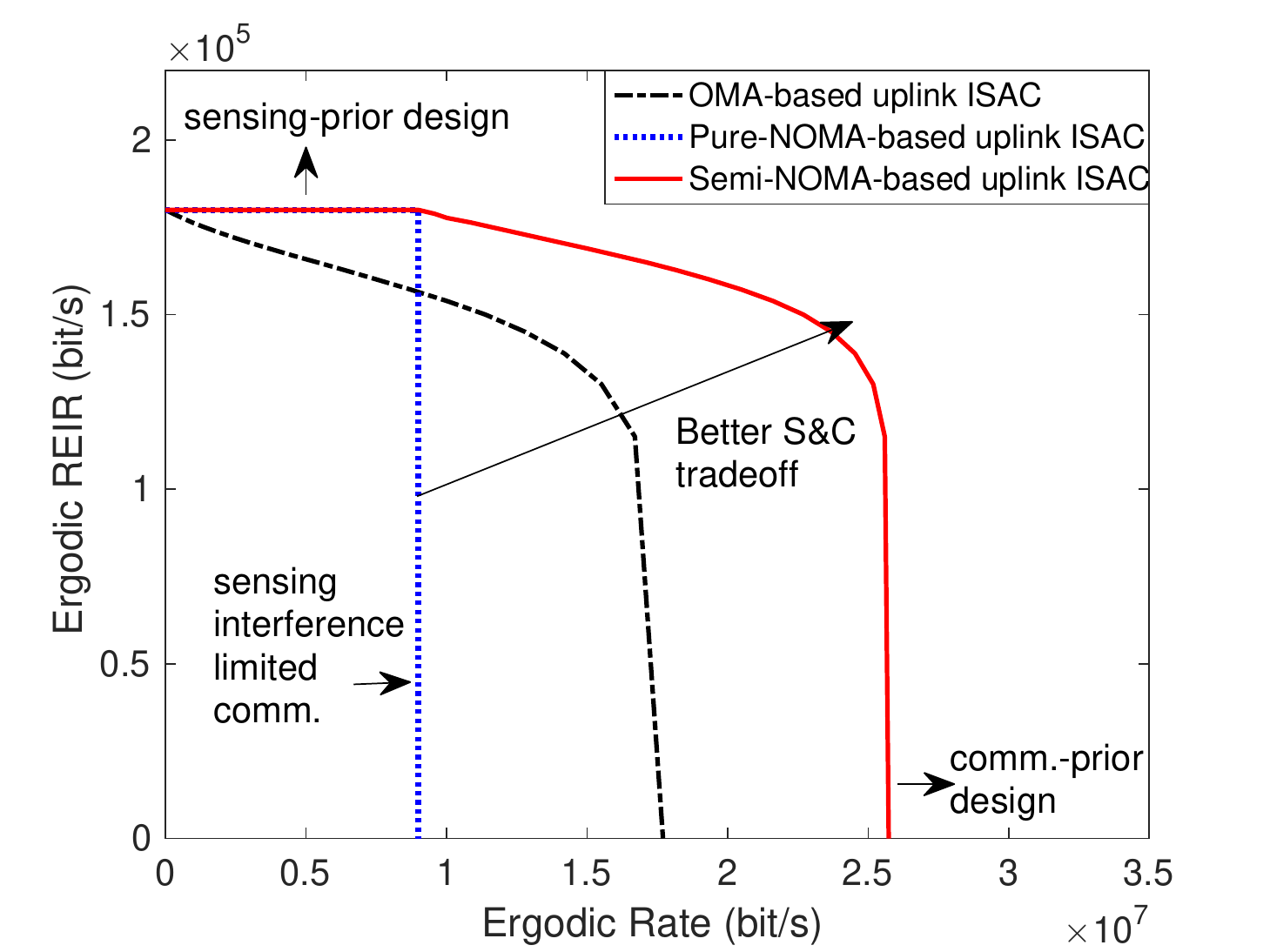}}
\caption{Uplink ISAC and initial numerical results. (a) An uplink ISAC system with OMA, pure-NOMA, and semi-NOMA schemes. (b) The achieved sensing-versus-communication tradeoff achieved by different uplink ISAC designs. The system parameters can be found in \cite[Section VI]{zhang2022semi}.}
\end{figure*}

Having discussed the employment of NOMA in the downlink ISAC, in this section, we focus our attention on the uplink ISAC. As discussed in the previous section, the key problem in the uplink ISAC is how to efficiently mitigate the mutual interference between the sensing echo and the communication signal that collide at the ISAC BS. Recall the fact that only the uplink communication signal conveys the information bits among the two types of signals. Based on this observation, we propose two designs for the uplink ISAC, namely 1) pure-NOMA based uplink ISAC and 2) semi-NOMA-based uplink ISAC. The key difference between the two designs is whether the radio resources are fully or partially shared between sensing and communication. In the following, we will introduce the principle as well as the advantages and disadvantages of each design.

\subsection{Pure-NOMA-based uplink ISAC}
As shown in Fig. \ref{fig:UL model}, let us take a basic uplink ISAC model as an example, where a ISAC BS aims to estimate the sensing-related parameters of one sensing target and recover the information messages uploaded by one communication user. In the conventional OMA-based uplink ISAC, the sensing echo and the communication signal can be allocated different orthogonal radio resources to facilitate each functionality in an interference-free manner, as illustrated in the top left of Fig. \ref{fig:UL model}. As discussed in Section II.A, the OMA-based uplink ISAC design may lead to low resource efficiency. To this end, a possible solution is to allow the sensing and communication functionalities to be realized via the fully shared radio resource, as illustrated in the top middle of Fig. \ref{fig:UL model}. For mitigating the inter-functionality interference caused by resource sharing, similar to the uplink NOMA principle employed in the uplink communication system, the ISAC BS will successively process the two types of signals with the aid of SIC. In contrast to the conventional NOMA uplink communication, where the SIC decoding order can be flexibility designed among the mixed communication signals~\cite{liu2022evolution}, the proposed pure-NOMA-based uplink ISAC design has to follow a fixed communication-to-sensing decoding order. This is because only the communication signal contains the information bits, which makes it possible to carry out SIC by first removing the communication signal from the received mixed sensing-communication signal. Then, the ISAC BS can analyze the sensing echo in an interference-free manner but have more available radio resources than the OMA-based uplink ISAC design. The procedure of mixed sensing-communication signal processing at the ISAC BS is illustrated at the bottom left of Fig. \ref{fig:UL model}.

Given the fixed communication-to-sensing decoding order, the communication functionality in the proposed pure-NOMA-based uplink ISAC design can be incorporated as an ``add-on'' component and has no negative effect on the existing sensing functionality. Therefore, on the one hand, the pure-NOMA-based uplink ISAC design is promising to be applied when the uplink ISAC has to be implemented over the spectrum that has been already occupied by sensing. On the other hand, observing that the sensing echo does not suffer from interference caused by the communication signal, the pure-NOMA-based uplink ISAC design is suitable for the sensing-prior design in ISAC, where the primary design objective is to guarantee the sensing performance. However, one main drawback of pure-NOMA-based uplink ISAC is that the fixed communication-to-sensing decoding order causes the uploaded information message to be always decoded against the sensing interference, thus resulting in a limited communication rate. Despite this limitation, the proposed pure-NOMA-based uplink ISAC design is practically implementable and beneficial. Note that, in practice, since the received sensing echo is originally transmitted by the ISAC BS (i.e., the sensing probing signal) and further reflected by the target, the signal strength of the received sensing echo might be lower than the communication signal due to the long-distance transmission. This nature of sensing echo promotes the successful implementation of SIC and an acceptable communication rate. It is also worth noting that even if only a limited communication rate can be achieved, the proposed pure-NOMA-based uplink ISAC design can be employed in IoT applications, where the uploading date rate required by IoT devices is usually low while the performance of sensing is prior. As a further advance, possible methods for sensing echo subtraction can be developed to first reduce the sensing interference caused to the communication. For example, the sensing echo can be predicted by exploiting the prior knowledge of the sensing target and other relevant sensing-related parameters. Based on the predicted sensing echo, the communication signal can be decoded with weakened sensing interference. This constitutes an interesting but challenging research topic for the proposed pure-NOMA-based uplink ISAC design.

\subsection{Semi-NOMA-based uplink ISAC Design}
In this subsection, we continue to propose a semi-NOMA-based uplink ISAC design, which not only unifies both the OMA-based and pure-NOMA-based uplink ISAC designs as special cases but also provides more flexible ISAC operations for well satisfying different objectives, e.g., sensing-prior design, communication-prior design, and sensing-versus-communication tradeoff design. As shown at the top right of Fig. \ref{fig:UL model}, the key idea behind the proposed semi-NOMA-based uplink ISAC design is to partition the total available radio resources into three orthogonal parts, namely the sensing-only resource block (S-RB), the communication-only RB (C-RB), and the mixed sensing-communication RB (S\&C-RB). On the one hand, in the S\&C-RB, the ISAC BS processes the received mixed sensing-communication signal following the principle of the pure-NOMA-based uplink ISAC design. On the other hand, the ISAC BS also processes the pure sensing echo/communication signal received from the S/C-RB in an interference-free manner as the OMA-based uplink ISAC design. Finally, the ISAC BS will combine the sensing/communication results obtained from the non-orthogonal S\&C-RB and the orthogonal S/C-RB together. From the perspective of resource allocation, since only partial radio resources are shared among the sensing and communication functionalities with the employment of NOMA (except in the two extreme cases), we therefore term it the \emph{semi-NOMA-based} uplink ISAC design.

The proposed semi-NOMA-based uplink ISAC design provides a general resource allocation framework for the uplink ISAC. By carefully optimizing the resource allocation among the three RBs, on the one hand, the semi-NOMA-based uplink ISAC design can be reduced to the OMA/pure-NOMA-based uplink ISAC design (e.g., make the S\&C-RB null or make both the S- and C-RBs null). On the other hand, it can further facilitate diversified uplink ISAC operations according to changes in ISAC design objectives, which cannot be realized by either the OMA-based or the pure-NOMA-based uplink ISAC design. It is also worth noting that the allocation of the C-RB in the proposed semi-NOMA-based uplink ISAC design is of vital importance in improving the communication performance since the communication signal received by the C-RB is not sensing-interference-limited compared to that in the pure-NOMA-based uplink ISAC design. As a result, a better sensing-versus-communication tradeoff can be achieved by the semi-NOMA-based uplink ISAC design. To further show the advantages of this concept, Fig. \ref{semi_result} compares the sensing-versus-communication tradeoff achieved by the OMA-based, pure-NOMA-based, and semi-NOMA-based uplink ISAC designs \cite{zhang2022semi}. Here, the ergodic radar estimation information rate (REIR) and ergodic communication rate are employed for evaluating the sensing and communication performance, respectively. As can be observed, the proposed semi-NOMA-based uplink ISAC achieves a superior sensing-versus-communication tradeoff than the other two designs. This result confirms the effectiveness of the proposed semi-NOMA-based uplink ISAC design. Moreover, it can be also found that the pure-NOMA-based ISAC is effective for the sensing-prior design but achieves the worst performance in the communication-prior design. This phenomenon is also consistent with our previous discussion. The superiority of the semi-NOMA-based uplink ISAC design mainly relies on flexible resource allocation between sensing and communication. This, however, results in numerous resource allocation variables that have to be optimized when the number of sensing targets and communication users is large, which is an important future research direction.

\section{Conclusions}
In this article, the development of ISAC has been investigated with NOMA from the MA perspective. Following the developing trend of non-orthogonal ISAC, the fundamental downlink and uplink ISAC models were introduced and the corresponding design challenges caused by the inter-functionality interference were also identified. To address these issues, two novel designs, namely NOMA-empowered downlink ISAC and NOMA-inspired downlink ISAC, were proposed for the downlink ISAC. By employing NOMA within communication and between the two functionalities, the two proposed designs are capable of effectively eliminating the typical inter-user interference and the new sensing-to-communication interference. Moreover, depending on whether radio resources are fully or partially shared by the two functionalities, a pure-NOMA-based uplink ISAC design and a semi-NOMA-based uplink ISAC design were proposed for the uplink ISAC. In particular, the pure-NOMA-based uplink ISAC design is promising to be employed in the sensing-prior design, while the semi-NOMA-based uplink ISAC design is capable of satisfying different ISAC design objectives given its provided flexible resource allocation. The provided numerical results confirmed the effectiveness of the proposed NOMA-ISAC designs. This article motivates several future research directions on NOMA-ISAC, including but not limited to the fundamental limit analysis, the resource allocation optimization, and the prototype implementation.

\bibliographystyle{IEEEtran}
\bibliography{mybib}

\end{document}